\documentclass[pre,reprint,amsmath,showpacs,showkeys,nofootinbib]{revtex4-1}

\usepackage{graphicx,amsfonts}
\usepackage{bbold}
\usepackage{hyperref,ifthen}

\begin{document}

\title{Reconstruction of Gaussian and log-normal fields with spectral smoothness}
\author{Niels Oppermann}\email{niels@mpa-garching.mpg.de}
\author{Marco Selig}
\author{Michael R. Bell}
\author{Torsten A. En{\ss}lin}
\affiliation{Max Planck Institute for Astrophysics, Karl-Schwarzschild-Str. 1, 85741 Garching, Germany}
\date{\today}


\begin{abstract}
    We develop a method to infer log-normal random fields from measurement data affected by Gaussian noise. The log-normal model is well suited to describe strictly positive signals with fluctuations whose amplitude varies over several orders of magnitude. We use the formalism of minimum Gibbs free energy to derive an algorithm that uses the signal's correlation structure to regularize the reconstruction. The correlation structure, described by the signal's power spectrum, is thereby reconstructed from the same data set. We show that the minimization of the Gibbs free energy, corresponding to a Gaussian approximation to the posterior marginalized over the power spectrum, is equivalent to the empirical Bayes ansatz, in which the power spectrum is fixed to its maximum a posteriori value. We further introduce a prior for the power spectrum that enforces spectral smoothness. The appropriateness of this prior in different scenarios is discussed and its effects on the reconstruction's results are demonstrated. We validate the performance of our reconstruction algorithm in a series of one- and two-dimensional test cases with varying degrees of non-linearity and different noise levels.
\end{abstract}

\maketitle

\section{Introduction}
\label{sec:introduction}

Reconstructing continuous fields from a finite and noisy data set is a problem encountered often in all branches of physics and the geo-sciences. In this paper we develop a method to reconstruct a log-normal field, i.e.\ a field whose logarithm can be modeled as a Gaussian random field, defined on an arbitrary manifold. We show simple one-dimensional examples as well as planar and spherical two-dimensional cases.

The log-normal model is well suited to describe many physical fields. Its main features are that the field values are guaranteed to be positive, that they may vary across many orders of magnitude, and that spatial correlations may exist. In cases in which the observations consist of counts of discrete objects, the data likelihood is often modeled as Poissonian. If the discrete objects are modeled to arise from an underlying log-normal field, such a model is known as a log Gaussian Cox process \citep{moller-1998} in the statistical literature. This model has been applied in fields as diverse as finance \citep{basu-2002}, agriculture \citep{brix-2001a}, atmospheric studies \citep{pati-2011}, and epidemiology \citep{brix-2001b,benes-2002}.

One prominent example from the discipline of astrophysics that is often modeled as a log-normal field is the matter density in today's universe \citep[see e.g.][]{coles-1991}. The simplest way to observationally estimate the matter density is to count the galaxies per volume. The relationship between the matter density and the galaxy number counts can be approximated as a Poisson process \citep[e.g.][]{kitaura-2008, kitaura-2010, jasche-2010b, jasche-2012}, thus making the data likelihood Poissonian as well with the expected number of galaxies per volume element given as a function of the underlying log-normal density field.

The main motivation for our work, however, comes from astronomical observations of radiative processes. The intensity of radiation coming from different directions in such observations can vary over many orders of magnitude while being spatially correlated. Thus applying a log-normal model for its statistical description seems natural. While photon counts bring with them a Poissonian likelihood, the observational uncertainty can be approximatively regarded to be Gaussian in the limit of large photon numbers. Therefore, we restrict ourselves to cases in which the measurement noise can be assumed to be additive and Gaussian.

Apart from the noise contribution, we will assume a deterministic linear relationship between the log-normal field and the data. This model is general enough to accommodate a large variety of observational settings, such as targeted point-wise observations across limited areas, convolutions of the log-normal field, interferometric observations leading to a Fourier transformed version of the log-normal field, or any combination of these effects.

The inclusion of spatial correlations in the field model is necessary for an accurate statistical description and will aid in the reconstruction. If the spatial correlation structure is known, knowledge of the field values at some locations can be used to extrapolate the field into regions where the field value has not been measured or the observations have a higher uncertainty. However, in general it is not known a priori how strongly the field is correlated across a given distance. In the case of a statistically homogeneous field, the correlation structure is described by the field's power spectrum. So if one wants to make use of the spatial correlations during the reconstruction, one needs to simultaneously infer the power spectrum. Several techniques have been developed to solve this problem for Gaussian fields \citep[e.g][]{wandelt-2004,jasche-2010a}. One such technique was derived within the framework of \textit{information field theory} \citep{ensslin-2009,lemm-2003} in \cite{ensslin_frommert-2011} and later rederived in \cite{ensslin_weig-2010}, where the formalism of minimum Gibbs free energy was introduced and employed to tackle this problem. In the same paper, the problem of inferring a log-normal field with unknown power spectrum in the presence of Poissonian noise was briefly discussed. \citet{kitaura-2012} use a Gibbs sampling procedure to reconstruct a log-normal field along with its power spectrum from Poissonian distributed data.

Here, we use the formalism of minimum Gibbs free energy to derive filter equations for the estimation of a log-normal field from data including a Gaussian noise component. The filter we present results in an estimate of the field, its power spectrum, and the respective uncertainties.

A problem that often arises when reconstructing power spectra is that the reconstructed power spectra do not turn out smooth, especially on scales with little redundancy, despite the true underlying spectra being smooth. One reason why this is not desirable is that sharp kinks and drop-offs in the power-spectrum can lead to ringing effects in the reconstructed field estimate. Another reason is that many of the fields occurring in nature are actually expected to have smoothly varying power over different scales since neighboring scales interact. One prominent example is the power spectrum of a viscous turbulent fluid \cite{kolmogorov-1941}. Smoothness of the power spectrum is often enforced by an ad-hoc smoothing step in the power spectrum reconstruction \cite{oh-1999,oppermann-2011a,oppermann-2012a}. Here, we follow an idea presented in \cite{ensslin_frommert-2011} and show how to enforce spectral smoothness by introducing an appropriate prior for the power spectrum that punishes non-smooth spectra. We demonstrate the feasibility of this approach using one specific example for such a smoothness prior for the reconstruction of both Gaussian and log-normal fields and discuss the range of applicability of the chosen spectral smoothness prior. Our approach also allows for the estimation of a complete uncertainty matrix for the estimated power spectrum.

We first develop the formalism of the spectral smoothness prior for the case of Gaussian signal fields in Sec.~\ref{sec:gaussian}, where we show how to derive the filter formulas of \cite{ensslin_frommert-2011} and \cite{ensslin_weig-2010} in a different way that easily accommodates an additional smoothness prior. After demonstrating the workings of the spectral smoothness prior, we use the Gibbs free energy formalism to derive filter formulas for the log-normal case and transfer the spectral smoothness results to this case in Sec.~\ref{sec:log-normal}. We demonstrate the performance of our log-normal reconstruction algorithm in a variety of test cases in Sec.~\ref{sec:lognormal-tests} and conclude in Sec.~\ref{sec:conclusions}.

\section{Reconstructing Gaussian fields with spectral smoothness}
\label{sec:gaussian}

First, we need to introduce some basic assumptions and notation. We mainly follow the notation that is used throughout the literature on \textit{information field theory} \citep[e.g.][]{ensslin-2009}.

Throughout the paper, we assume that we are analyzing a set of data $d = \left(d_1,\dots,d_r\right) \in \mathbb{R}^r$ that depends on an underlying signal field $s:~\mathcal{M}\longrightarrow \mathbb{R}$, subject to additive noise $n = \left(n_1,\dots,n_r\right) \in \mathbb{R}^r$,
\begin{equation}
    d = f(s) + n.
\end{equation}
Here, $\mathcal{M}$ is the discrete or continuous space or manifold that the signal is defined on. In Sec.~\ref{sec:log-normal} we will discuss the examples $\mathcal{M} = \mathcal{S}^1$, $\mathcal{S}^2$, and $\mathcal{T}^2$. In this section, we assume the relationship between signal field and data to be linear, described by a response operator $R$, so that
\begin{equation}
    \label{eq:data-model-linear}
    d = Rs + n.
\end{equation}
In most applications, the response operator will include some instrumental or observational effects such as observations in specific locations of $\mathcal{M}$, convolutions of $s$ with an instrumental response function, or a Fourier transform of the signal field. The only restriction that we make here is that the operation that generates the data from the signal has to be linear in the signal, which is a good approximation in the astrophysical applications that we envision and also in many other contexts. In Sec.~\ref{sec:log-normal} we present a case with a non-linear response function that arises from the incorporation of non-Gaussian features of the signal. Finally, we restrict ourselves to real-valued signals only in the interest of notational simplicity. All our results can be straightforwardly generalized to complex-valued signal fields.

For the noise $n$, we assume Gaussian statistics,
\begin{equation}
    n \hookleftarrow \mathcal{G}(n,N),
\end{equation}
    described by a not necessarily diagonal covariance matrix $N$. Here, $\mathcal{G}(\phi,\Phi)$ denotes a multi-variate Gaussian probability distribution function with covariance $\Phi$,
\begin{equation}
    \mathcal{G}(\phi,\Phi) = \frac{1}{\left| 2\pi \Phi \right|^{1/2}} \exp\left(-\frac{1}{2} \phi^\dagger \Phi^{-1} \phi \right).
\end{equation}
We use the $\dagger$-symbol to denote a transposed (and in general complex conjugated) quantity and take the limit of an infinite-dimensional Gaussian whenever the argument is a continuous field, with the appropriate limit of the scalar product
\begin{equation}
    \phi^\dagger \psi = \int_\mathcal{M} \!\!\! {\mathrm{d}x\,\overline{\phi(x)}\, \psi(x)}~~\forall~\phi,\psi:~ \mathcal{M}\longrightarrow \mathbb{R}.
\end{equation}

In this section we will deal with the case that the signal field $s$ can also be regarded -- or at least approximated -- as a zero-mean Gaussian random field with covariance $S$,
\begin{equation}
    s \hookleftarrow \mathcal{G}(s,S).
\end{equation}
It can be straightforwardly shown \citep[e.g.][]{ensslin-2009} that the posterior mean $m$ of the signal field under these assumptions is given by
\begin{equation}
    \label{eq:WF}
    m = \int \!\! \mathcal{D}s\, s\, \mathcal{P}(s|d) = Dj.
\end{equation}
Here, $\int\mathcal{D}s$ denotes an integral over the configuration space of all possible signal realizations. The operator
\begin{equation}
    \label{eq:D}
    D = \left(S^{-1} + R^\dagger N^{-1} R\right)^{-1}
\end{equation}
is known as the \textit{information propagator} and the field
\begin{equation}
    \label{eq:j}
    j = R^\dagger N^{-1} d
\end{equation}
is called \textit{information source}.

In these formulas, the presence of the Gaussian signal prior described by the signal covariance serves as a means of regularization of the desired continuous signal field reconstruction which otherwise is under-determined when only constrained by the finite and noisy data set. However, in most physical applications, the signal covariance is not known a priori. The problem of reconstructing Gaussian fields with unknown signal covariances has in principle been solved \cite{ensslin_frommert-2011, ensslin_weig-2010, wandelt-2004, jasche-2010a}, and even an unknown noise covariance can be overcome \cite{oppermann-2011b}.

En{\ss}lin~\&~Weig~\cite{ensslin_weig-2010} use the formalism of minimum Gibbs free energy to derive filter formulas to be applied to the data set when the signal's covariance is unknown. We will review this formalism briefly in Sec.~\ref{sec:log-normal}, where we employ it to reconstruct log-normal signal fields.

Under the assumption of statistical homogeneity and isotropy, the unknown signal covariance becomes diagonal in the harmonic eigenbasis, i.e.\ the Fourier basis for signals defined on Euclidean space and the spherical harmonics basis for signals defined on the sphere. In the following, we denote as $\vec{k}$ the vector of parameters determining one mode in the harmonic decomposition, i.e.\ $\vec{k}=(k_1,\dots,k_n)$ for $n$-dimensional Euclidean space and $\vec{k}=(\ell,m)$ for the two-sphere, where $\ell$ is the angular momentum quantum number and $m$ the azimuthal one. Furthermore, $k$ shall stand for the scale of the harmonic component, i.e.\ $k=\sqrt{k_1^2 + \cdots + k_n^2}$ and $k=\ell$ for $\mathbb{R}^n$ and $\mathcal{S}^2$, respectively.

Due to the isotropy assumption, the diagonal of the signal covariance matrix, the signal's power spectrum $P_k$, depends only on the scale $k$ and not on the specific mode $\vec{k}$, i.e.
\begin{equation}
    \label{eq:S-homo-iso}
    S_{\vec{k}\vec{k}'} = \delta_{\vec{k}\vec{k}'} P_k.
\end{equation}
Under these symmetry assumptions, the filter formulas derived by En{\ss}lin~\&~Weig~\cite{ensslin_weig-2010} via the minimization of the Gibbs free energy can be written as
\begin{align}
    \label{eq:critical-m}
    m &= Dj,\\
    \label{eq:critical-P}
    P_k &= \frac{1}{\alpha_k - 1 + \frac{\rho_k}{2}} \left(q_k + \frac{1}{2} \sum_{\left\{\vec{k}' | k' = k\right\}} \!\!\!\!\! \left(\left|m_{\vec{k}'}\right|^2 + D_{\vec{k}'\vec{k}'} \right) \right).
\end{align}
Note that the first of these equations is the same as in the case of a known signal covariance, Eq.~\eqref{eq:WF}, a generalized Wiener filter \cite{ensslin-2009}. Here, $\alpha_k$ and $q_k$ are parameters used to determine the priors for the spectral coefficients $P_k$ (see Sec.~\ref{sec:log-normal} for details) and
\begin{equation}
    \rho_k = \sum_{\left\{\vec{k}' | k' = k\right\}} \!\!\!\!\! 1
\end{equation}
is the number of degrees of freedom on the scale $k$. We will show in the next subsection that these formulas can also be regarded as the result of an empirical Bayes ansatz, in which the unknown power spectrum is replaced by its maximum a posteriori estimate.

This formalism has been successfully applied for astrophysical reconstructions \cite{oppermann-2011a,oppermann-2012a} and in a more abstract computational setting \cite{selig-2012}. In these applications, Eqs.~\eqref{eq:critical-m} and \eqref{eq:critical-P} were simply iterated. However, they were supplemented with an additional ad-hoc smoothing step for the power spectrum as part of each iteration.

\begin{figure}
    \input{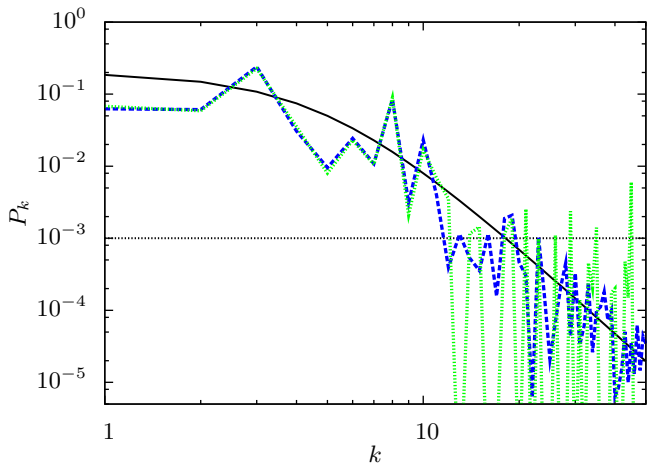}
    \caption{\label{fig:powspec-nosmooth}Power spectra of the one-dimensional test-case. The black solid line shows the theoretical power spectrum, the blue dashed line the power in the randomly drawn signal realization studied here, and the green dotted line shows the power spectrum reconstructed without smoothing. The horizontal dotted line indicates the noise power, given by $\sigma_n^2/100$.}
\end{figure}

\begin{figure}
    \input{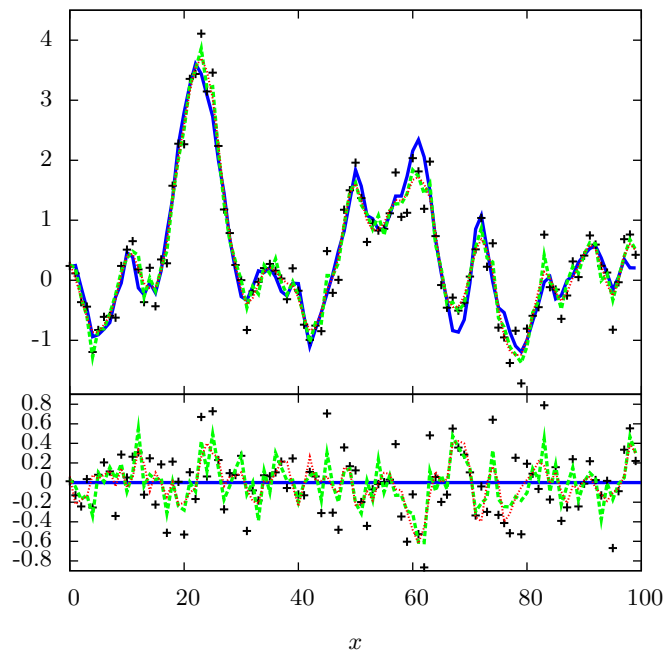}
    \caption{\label{fig:map-nosmooth}Signal reconstruction in the one-dimensional test-case. The blue solid line shows the randomly drawn signal realization, the crosses are noisy data points used to do the reconstruction, the green dashed line shows the signal reconstruction obtained without spectral smoothing and the red dotted line is the Wiener filter reconstruction. The lower panel shows the same with the true signal substracted.}
\end{figure}

To illustrate the usefulness of a smoothing step, we set up a simple one-dimensional test-case. Here, we assume that our signal field is defined on the one-sphere $\mathcal{S}^1$, i.e.\ the interval $[0,1)$ with periodic boundary conditions, which we discretize into $100$ pixels. We make up a power spectrum for the signal of the form
\begin{equation}
    \label{eq:powspec}
    P_k = P_0 \left(1 + \left(\frac{k}{k_0}\right)^2\right)^{-\gamma/2},
\end{equation}
where we choose $P_0 = 0.2$, $k_0 = 5$, and $\gamma = 4$. Figures~\ref{fig:powspec-nosmooth} and \ref{fig:map-nosmooth} show the power spectrum and the signal realization randomly drawn from it, respectively.

Furthermore, we assume that we have measured the signal field in each of the $100$ pixels once, subject to homogeneous and uncorrelated Gaussian noise with variance $\sigma_n^2 = 0.1$. In the formalism of Eq.~\eqref{eq:data-model-linear}, this corresponds to $R = \mathbb{1}$ and $N = \sigma_n^2 \mathbb{1}$. The data realization is also shown in Fig.~\ref{fig:map-nosmooth}.

We plot in Figs.~\ref{fig:powspec-nosmooth} and \ref{fig:map-nosmooth} the power spectrum and map reconstruction, respectively, obtained by iterating Eqs.~\eqref{eq:critical-m} and \eqref{eq:critical-P} without smoothing. While the signal reconstruction suffers from some over-estimation of small-scale fluctuations, the reconstructed power spectrum fluctuates wildly for $k > 10$ and is therefore not trustworthy at all on these scales. The scales on which the reconstructed power by chance peaks above the noise level are the ones that are especially misrepresented in the reconstructed signal field.

This example should serve to illustrate that some kind of spectral smoothing is necessary, especially in cases where one is interested in the power spectrum itself. In Fig.~\ref{fig:map-nosmooth} we also show a comparison to the Wiener filter reconstruction, i.e.\ Eq.~\eqref{eq:WF}, using the true power spectrum. In the bottom panel of Fig.~\ref{fig:map-nosmooth} it can be seen that the residuals, i.e.\ the differences between the true signal and its reconstruction, are reduced if a power spectrum is used that is closer to the true one. Smoothness is one aspect of the true power spectrum that can be used to constrain its reconstruction.

The two-point correlation function of the signal provides another way of looking at the smoothness property of its power spectrum. For a statistically homogeneous signal, the power spectrum is the Fourier transform of the correlation function and vice versa. Thus, a power spectrum that exhibits fluctuations on arbitrarily small scales in $k$-space corresponds to a signal that exhibits correlations over arbitrarily large scales in position space. Turning this argument around, any signal with correlations only over a finite range in position space or at least correlations that are decreasing rapidly with distance will have a power spectrum that does not exhibit any features on arbitrarily small scales in $k$-space. Thus, the power spectrum can be expected to be smooth on the reciprocal scale of the typical correlation length scale. 

In the remainder of this section, we will show how to incorporate a prior enforcing spectral smoothness into the formalism presented thus far.

\subsection{Reconstruction as a combination of \textit{posterior mean} and \textit{maximum a posteriori}}
\label{sec:combi-derivation}

Before incorporating a spectral smoothness prior into the derivation of Eqs.~\eqref{eq:critical-m} and \eqref{eq:critical-P}, we will rederive them in a way different from the one presented in \cite{ensslin_frommert-2011,ensslin_weig-2010}.

As was already mentioned, Eq.~\eqref{eq:critical-m} corresponds to the posterior mean of the signal under the assumption of a known covariance matrix, i.e.\ a known power spectrum. Since the posterior probability distribution is Gaussian in this case, its mean also maximizes the probability, so that $m$ is the posterior mean and the maximum a posteriori solution at the same time,
\begin{equation}
    m = \int \!\! \mathcal{D}s\, s\, \mathcal{P}(s|d,P) = \underset{s}{\mathrm{argmax}} \left\{\mathcal{P}(s|d,P)\right\}.
\end{equation}

Here, we will show that Eq.~\eqref{eq:critical-P} can be written as a maximum a posteriori solution as well, considering the signal-marginalized posterior
\begin{equation}
    \mathcal{P}(P|d) = \int \!\! \mathcal{D}s\,\mathcal{P}(s,P|d).
\end{equation}
In order to do this, we first have to define a prior for the power spectrum $P$. In accordance with \cite{ensslin_frommert-2011,ensslin_weig-2010}, we choose independent inverse-gamma distributions for each spectral component $P_k$,
\begin{align}
    \mathcal{P}(P) &= \prod_k \mathcal{P}_\mathrm{IG}(P_k)\nonumber\\
    \label{eq:inverse-gamma}
    &= \prod_k \frac{1}{q_k \Gamma(\alpha_k - 1)} \left(\frac{P_k}{q_k}\right)^{-\alpha_k} \exp\left(-\frac{q_k}{P_k}\right).
\end{align}
Here, $\Gamma(\cdot)$ denotes the gamma function, $q_k$ is a parameter defining the location of an exponential cut-off at low values of $P_k$, and $\alpha_k$ defines the slope of a power-law for large values of $P_k$. By tuning these parameters, the prior can be narrowed or widened according to the a priori knowledge about the power spectrum. Taking the limit $q_k \rightarrow 0$ and $\alpha_k \rightarrow 1$ turns the inverse-gamma distribution into Jeffreys prior which is flat on a logarithmic scale. In the examples presented in this paper we always take this limit in the final filter formulas.

In the following, we will work with the logarithmic spectral components
\begin{equation}
    p_k = \log P_k.
\end{equation}
The corresponding prior for these can be straightforwardly derived from the conservation of probability under variable transformations and reads
\begin{align}
    \mathcal{P}(p) &= \mathcal{P}(P)\, \left|\frac{\mathrm{d}P}{\mathrm{d}p}\right|\nonumber\\
    &= \prod_k \frac{q_k^{\alpha_k - 1}}{\Gamma(\alpha_k - 1)} \mathrm{e}^{-\left(\alpha_k - 1\right) p_k} \exp \left(-q_k \mathrm{e}^{-p_k}\right).
\end{align}
Using this prior, we can calculate the signal-marginalized probability of data and power spectrum, $\mathcal{P}(d,p)$, and its negative logarithm, the Hamiltonian
\begin{align}
    H(d,p) &= -\log \mathcal{P}(d,p)\nonumber\\
    &= -\log \int \!\! \mathcal{D}s\, \mathcal{G}(d-Rs,N)\, \mathcal{G}(s,S)\, \mathcal{P}(p)\nonumber\\
    &= \frac{1}{2} \mathrm{tr} \left(\log S\right) - \frac{1}{2} \mathrm{tr} \left(\log D\right) - \frac{1}{2} j^\dagger D j\nonumber\\
    \label{eq:H-nosmooth}
    &~~~ + \sum_k \left(\left(\alpha_k - 1\right) p_k + q_k \mathrm{e}^{-p_k}\right) + \mathrm{const.},
\end{align}
where we have made use of the definitions \eqref{eq:D} and \eqref{eq:j} and have collected all terms that do not depend on the power spectrum into an unimportant additive constant. Using the spectral dependence of the signal covariance matrix in a statistically homogeneous and isotropic setting, Eq.~\eqref{eq:S-homo-iso}, we can take the derivative of the Hamiltonian with respect to one $p_k$ and equate it to zero, thus maximizing the posterior probability of the logarithmic power spectrum. The resulting equation is exactly Eq.~\eqref{eq:critical-P} for $P_k = \mathrm{e}^{p_k}$ if one makes the identification $m=Dj$.

Thus we have shown that the filter formulas, Eqs.~\eqref{eq:critical-m} and \eqref{eq:critical-P}, can be derived as a combination of posterior mean for the signal reconstruction and maximum a posteriori for the power spectrum. This effectively means that we have made the approximation
\begin{align}
    m &= \int \!\! \mathcal{D}s\, s\, \mathcal{P}(s|d)\nonumber\\
    &= \int \!\! \mathcal{D}s\, \!\! \int \!\! \mathcal{D}p\, s\, \mathcal{P}(s|d,p) \mathcal{P}(p|d)\nonumber\\
    &\approx \int \!\! \mathcal{D}s\, \!\! \int \!\! \mathcal{D}p\, s\, \mathcal{P}(s|d,p)\, \delta{\left(p - p^{\mathrm{(MAP)}}\right)},
\end{align}
i.e.\ we have approximated the posterior probability distribution for the power spectrum with a delta distribution centered on its maximum, a procedure known as empirical Bayes method \citep[e.g.][]{robbins-1955}.

It may be worth noting that in the formalism of the maximum a posteriori solution for the power spectrum, it is straightforward to derive a rough uncertainty estimate as well. The Hessian of the Hamiltonian gives the curvature of the posterior probability distribution and its inverse can thus be regarded as an uncertainty matrix. For the Hamiltonian given in Eq.~\eqref{eq:H-nosmooth} we obtain
\begin{align}
    & \left.\frac{\partial^2 H(d,p)}{\partial p_k \partial p_{k'}}\right|_{p=p^{\mathrm{(MAP)}}} = \left(\alpha_k - 1 + \frac{\rho_k}{2}\right) \delta_{kk'}\nonumber\\
    &\quad \left. - \frac{1}{2} \mathrm{e}^{-p_k - p_{k'}} \!\!\!\!\sum_{{\left\{\vec{q} | q = k\right\}} \atop {\left\{\vec{q}' | q' = k'\right\}}}\!\!\!\!\! \left( 2\, \Re{\left(m_{\vec{q}} m_{\vec{q}'}^* D_{\vec{q}\vec{q}'}\right)} + \left|D_{\vec{q}\vec{q}'}\right|^2 \right) \right|_{p=p^{\mathrm{(MAP)}}},
\end{align}
where $\Re{(\cdot)}$ denotes the real part of a complex number.

\subsection{Spectral smoothness priors}
\label{sec:smoothness-prior}

Here, we show how to incorporate a spectral smoothness prior into the formalism developed in the previous section. We do this by augmenting the inverse-gamma distributions previously assumed as the spectral prior with a probability distribution that enforces smoothness of the power spectrum, so that
\begin{equation}
    \mathcal{P}(p) = \mathcal{P}_\mathrm{sm}(p) \prod_k \mathcal{P}_\mathrm{IG}(p_k).
\end{equation}
As an example, we choose a smoothness-enforcing prior of the shape
\begin{equation}
    \label{eq:smoothnessprior}
    \mathcal{P}_\mathrm{sm}(p) \propto \exp \left( -\frac{1}{2 \sigma_p^2} \int \!\! \mathrm{d}{\left(\log k\right)}\, \left(\frac{\partial^2 \log P_k}{\partial \left(\log k\right)^2}\right)^2 \right).
\end{equation}
This prior is maximized by any power-law power spectrum and punishes deviations from such a shape. The strength of the punishment is determined by the parameter $\sigma_p$. Other useful shapes for a smoothness prior could contain the first logarithmic derivative, punishing steep spectra, or simple derivatives with respect to $k$, punishing abrupt changes in the power spectrum. To illustrate the meaning of such smoothness priors and point out possible caveats, we discuss a few specific cases in App.~\ref{app:casestudies}.

The spectral smoothness prior, Eq.~\eqref{eq:smoothnessprior}, can be written as a Gaussian in $p = \log P$,
\begin{equation}
    \label{eq:smoothnesspriorwithT}
    \mathcal{P}_\mathrm{sm}(p) \propto \exp \left( -\frac{1}{2} p^\dagger T p \right),
\end{equation}
where the linear operator $T$ includes both the second derivative and the scaling constant $\sigma_p^2$. The detailed form of the operator $T$ that we use in our calculations can be found in App.~\ref{app:T-matrix}.

Introducing this prior corresponds to adding the term $\frac{1}{2} p^\dagger T p$ to the Hamiltonian in Eq.~\eqref{eq:H-nosmooth}. Taking its derivative with respect to one $p_k$ and equating the result with zero, we now get
\begin{equation}
    \label{eq:critical-P-withsmooth}
    \mathrm{e}^{p_k} = \frac{q_k + \frac{1}{2} \sum_{\left\{\vec{k}' | k' = k\right\}} \left(\left|m_{\vec{k}'}\right|^2 + D_{\vec{k}'\vec{k}'} \right)}{\alpha_k - 1 + \frac{\rho_k}{2} + \left(Tp\right)_k}.
\end{equation}
The only point in which this equation differs from Eq.~\eqref{eq:critical-P} is the extra term $Tp$ in the denominator. As before, we can calculate the inverse Hessian of the Hamiltonian as an approximate uncertainty matrix. The Hessian is
\begin{align}
    \label{eq:Hessian-smooth}
    & \left.\frac{\partial^2 H(d,p)}{\partial p_k \partial p_{k'}}\right|_{p=p^{\mathrm{(MAP)}}} = \left(\alpha_k - 1 + \frac{\rho_k}{2} + \left(Tp\right)_k\right) \delta_{kk'} + T_{kk'}\nonumber\\
    &\quad \left. - \frac{1}{2} \mathrm{e}^{-p_k - p_{k'}} \!\!\!\! \sum_{{\left\{\vec{q} | q = k\right\}} \atop {\left\{\vec{q}' | q' = k'\right\}}} \!\!\!\!\! \left( 2\, \Re{\left(m_{\vec{q}} m_{\vec{q}'}^* D_{\vec{q}\vec{q}'}\right)} + \left|D_{\vec{q}\vec{q}'}\right|^2 \right) \right|_{p=p^{\mathrm{(MAP)}}}.
\end{align}

\subsection{Test-cases}
\label{sec:gaussian-test-cases}

\begin{figure}
    \input{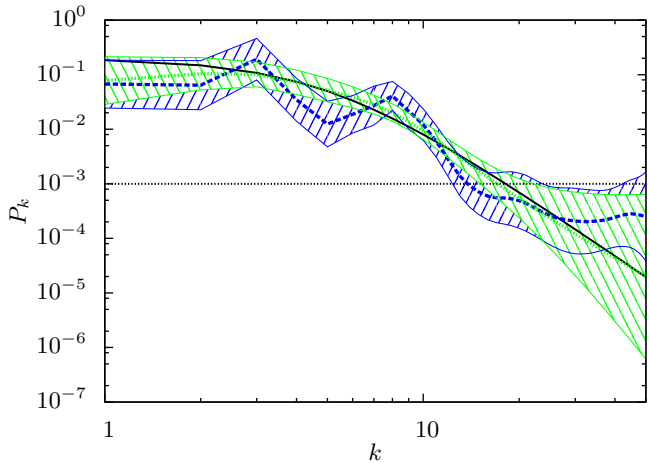}
    \caption{\label{fig:powspec-smooth}Power spectra of the one-dimensional test-case with smoothness prior. The black solid line shows the theoretical power spectrum, the blue dashed line the power spectrum reconstructed with a smoothness prior with $\sigma_p^2 = 1000$, and the green dotted line the reconstruction with a stiffer smoothness prior with $\sigma_p^2 = 10$. The hatched regions around the dashed and dotted lines are the corresponding one-sigma uncertainty regions estimated from the inverse Hessian of the Hamiltonian. The horizontal dotted line indicates the noise level.}
\end{figure}

Using the same one-dimensional test-case as shown in Figs.~\ref{fig:powspec-nosmooth} and \ref{fig:map-nosmooth}, we perform a reconstruction using the spectral smoothness prior given in Eq.~\eqref{eq:smoothnessprior}. The resulting power spectra using a moderate strength for the prior with $\sigma_p^2 = 1000$ and a strict smoothness prior with $\sigma_p^2 = 10$ are shown in Fig.~\ref{fig:powspec-smooth}. Clearly, the new reconstructions are a much better approximation to the shape of the theoretical power spectrum than the one shown in Fig.~\ref{fig:powspec-nosmooth}. Also, as expected, the power spectrum obtained when using the strict smoothness prior turns out smoother and more closely resembles a power law.

\begin{figure}
    \begin{center}
        \includegraphics[width=1.0\columnwidth]{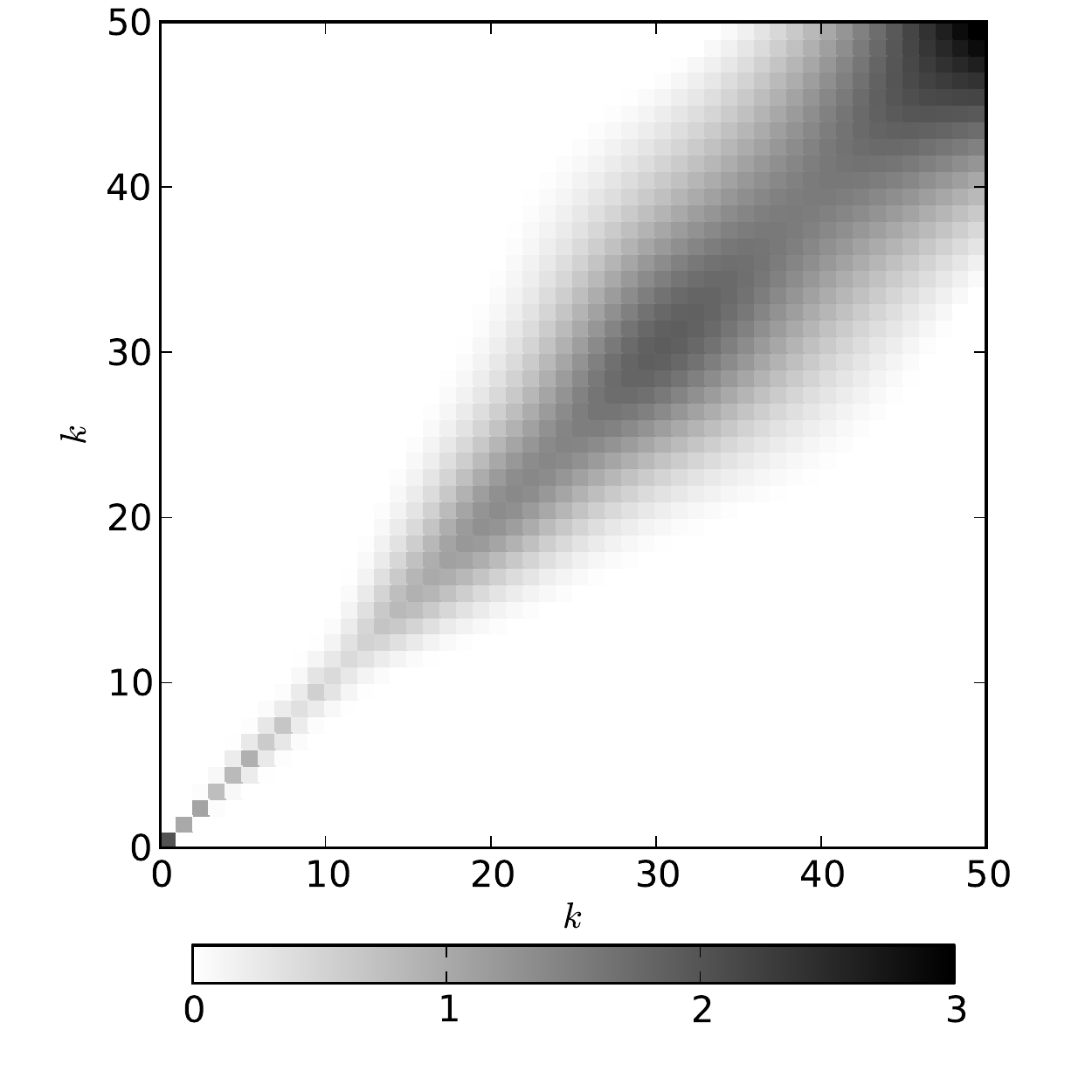}
        \caption{\label{fig:matrix}Inverse Hessian of the Hamiltonian with a smoothness prior with $\sigma_p^2 = 1000$. Plotted is the full matrix given by the inverse of Eq.~\eqref{eq:Hessian-smooth}, evaluated at $p = p^{\mathrm{(MAP)}}$. The diagonal of this matrix can be translated into the uncertainty interval plotted in Fig.~\ref{fig:powspec-smooth}, however, that does not take into account the correlations across different $k$-modes that are visible in this plot.}
    \end{center}
\end{figure}

Also shown in Fig.~\ref{fig:powspec-smooth} is an uncertainty estimate for the reconstructed power spectra, obtained from the Hessian of the Hamiltonian. Taking a one-sigma uncertainty estimate for the logarithmic power spectrum components,
\begin{equation}
    \Delta_p = \left[\mathrm{diag}\left(\left(\left.\frac{\delta^2 H(d,p)}{\delta p \, \delta p^\dagger}\right|_{p = p^{\mathrm{(MAP)}}}\right)^{-1}\right)\right]^{1/2},
\end{equation}
we plot the uncertainty interval for the power spectra as $P_k = \exp\left(p_k^{\mathrm{(MAP)}} \pm \Delta_p\right)$. The uncertainty interval of the power spectrum estimates calculated in this way can, however, be misleading. It should be noted that the uncertainty of the power on the different $k$-modes is correlated. To illustrate this, we plot the complete uncertainty matrix, i.e.\ the inverse Hessian of the Hamiltonian, for the case with $\sigma_p^2 = 1000$ in Fig.~\ref{fig:matrix}. It can be seen from this figure that the correlation of the uncertainty on different $k$-modes is especially large for small scales. Furthermore, the inverse Hessian is only a rough approximation to the uncertainties in the power spectrum estimation, which typically underestimates these.

In all following examples we will use a spectral smoothness prior with an intermediate stiffness, given by $\sigma_p^2 = 100$.

\begin{figure}
    \input{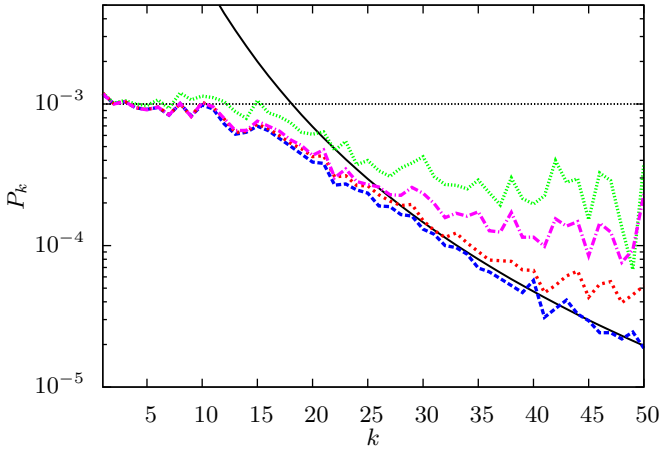}
    \caption{\label{fig:diffpower}Residual power between true signals and their reconstructions averaged over $100$ different realizations. The black solid line shows the theoretical power spectrum of the signals, given by Eq.~\eqref{eq:powspec} with $P_0 = 0.2$, $k_0 = 5$, and $\gamma = 4$. The dotted horizontal line indicates the noise level, corresponding to $\sigma_n^2 = 0.1$. The remaining lines show the residual power for different reconstruction schemes. From top to bottom, these are a reconstruction without any spectral smoothing (green dotted line), with an ad-hoc convolution of the power spectrum (magenta dash-dotted line), with the smoothness prior given by Eq.~\eqref{eq:smoothnessprior} with $\sigma_p^2 = 100$ (red short-dashed line), and using the correct power spectrum (blue long-dashed line).}
\end{figure}

\begin{figure}
    \input{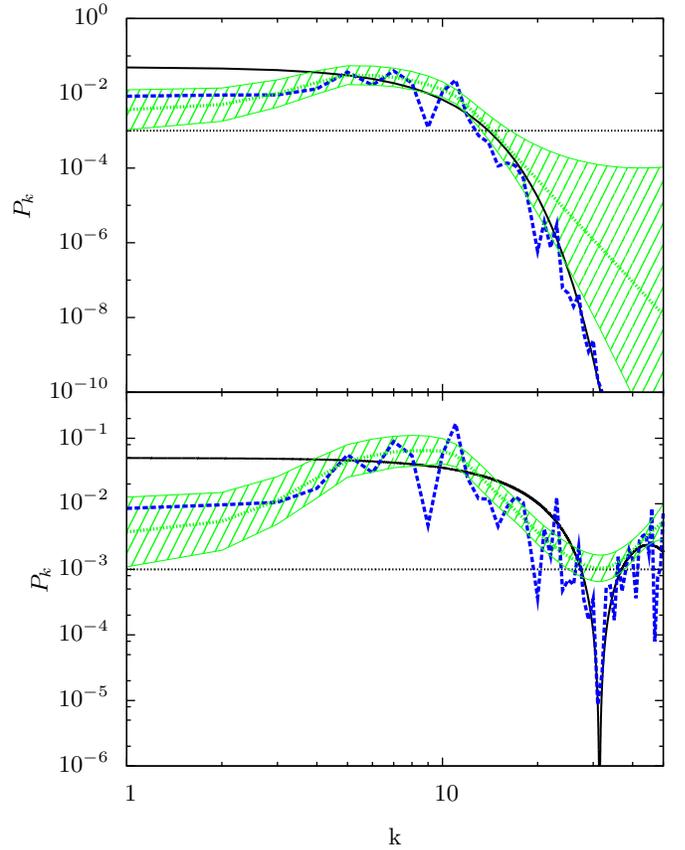}
    \caption{\label{fig:powspec-problematic}Power spectra of signals with correlation functions given by Eq.~\eqref{eq:corr-gauss} (top panel) and Eq.~\eqref{eq:corr-ricehat} (bottom panel). The black solid lines show the theoretical power spectra, the blue dashed lines the power in the random field realization drawn from the theoretical power spectra, and the green dotted lines the reconstructed power spectra using the spectral smoothness prior given in Eq.~\eqref{eq:smoothnessprior} with $\sigma_p^2 = 100$. The parameters substituted into Eq.~\eqref{eq:corr-gauss} are $\sigma = 0.2$ and $C_0 = 1/(4\sqrt{2\pi})$ and the ones used in Eq.~\eqref{eq:corr-ricehat} are $L = 0.2$ and $C_0 = 0.25$. The noise variance in both cases is $\sigma_n^2 = 0.1$. The hatched area indicates the uncertainty of the reconstructed power spectrum estimated from the inverse Hessian and the horizontal dotted line the noise level.}
\end{figure}

\begin{figure}[!]
    \input{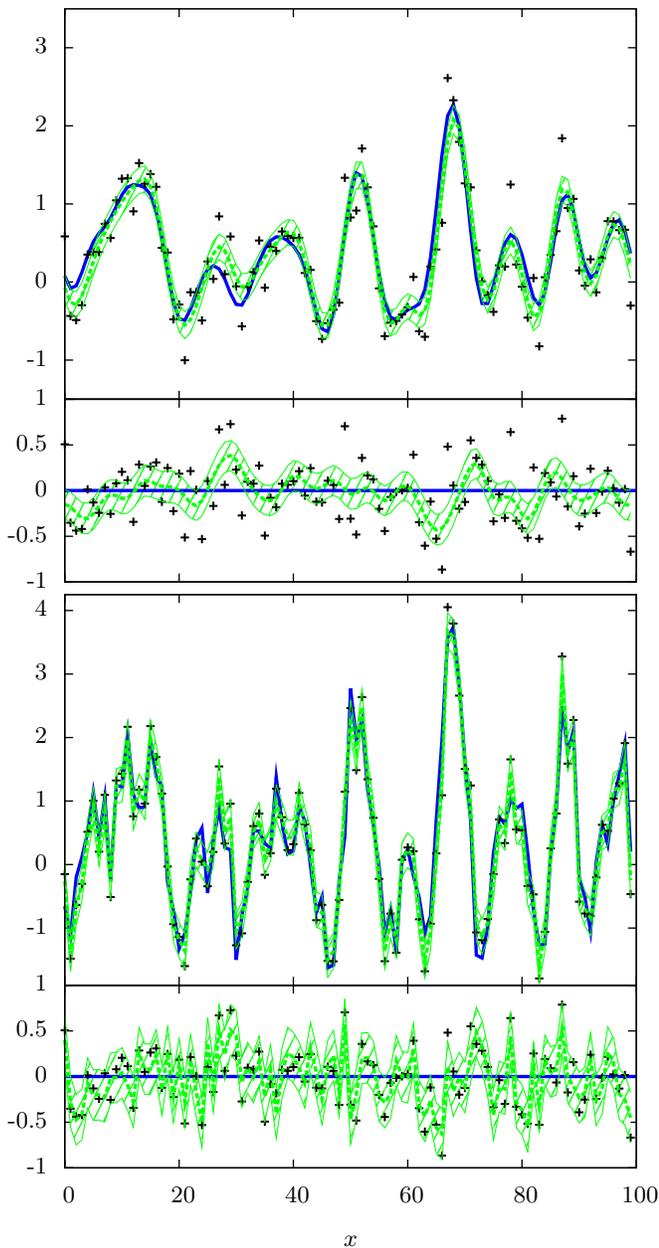}
    \caption{\label{fig:maps-problematic}Signal fields corresponding to the power spectra shown in Fig.~\ref{fig:powspec-problematic}. The top panel shows a signal with a Gaussian two-point correlation function, Eq.~\eqref{eq:corr-gauss}, the bottom panel shows one with a triangular correlation function, Eq.~\eqref{eq:corr-ricehat}. Shown in each plot are the signal realization $s$ (blue solid line), the data $d$ (crosses), the signal reconstruction $m$ (green dashed line), and an estimate of the local one-sigma uncertainty of the reconstruction, given by $m \pm \mathrm{diag}(D)^{1/2}$ (hatched area). At the bottom of each panel, the same is plotted with the true signal subtracted.}
\end{figure}

To study the improvement that the spectral smoothness prior brings for the reconstruction of the signal field, we draw $100$ different signal realizations from the same power spectrum and add $100$ different noise realizations. For each of these data sets, we perform the full reconstruction and then calculate the power of the difference between the reconstructed field $m$ and the true signal field $s$. In Fig.~\ref{fig:diffpower} we plot this power, averaged over the $100$ realizations, for different reconstruction schemes. Under the assumption that the correct power spectrum is known a priori, the residual power is essentially given by the noise level on scales for which the signal is dominating and by the signal power on scales for which the noise is dominating. As can be seen from the plot, reconstructing the power spectrum without any spectral smoothing leads to a significantly increased residual power on the noise-dominated scales, while the quality of the reconstructions including the smoothness prior is close to the one of the Wiener filter reconstructions, showing only a slight excess in residual power on the smallest scales. We also plot the average residual power for reconstructions in which an ad-hoc smoothing step is applied to the power spectrum after each iteration of Eq.~\eqref{eq:critical-P}. This is implemented as a simple convolution with a Blackman window of width $\Delta_k = 9$. As can be seen in Fig.~\ref{fig:diffpower}, this ad-hoc smoothing partly alleviates the problems of the power spectrum reconstruction but is clearly outperformed by the rigorous application of a smoothness prior.

Finally, we consider signal fields of the types discussed in App.~\ref{app:Gaussian} and \ref{app:ricehat}, i.e.\ signals with Gaussian and triangular correlation functions given by Eqs.~\eqref{eq:corr-gauss} and \eqref{eq:corr-ricehat}, respectively. As discussed in the appendix, these correlation functions lead to theoretical power spectra that are strongly suppressed by our spectral smoothness prior. Here, we investigate how serious this unwanted effect is in practice.

Shown in Figs.~\ref{fig:powspec-problematic} and \ref{fig:maps-problematic} are the results of the reconstruction for the power spectra and signal fields, respectively. We again use signals defined on an interval of length one, which we divide into $100$ pixels, and choose $\sigma = 0.2$ and $C_0 = 1/(4\sqrt{2\pi})$ for the Gaussian correlation function, Eq.~\eqref{eq:corr-gauss}, and $L = 0.2$ and $C_0 = 0.25$ for the triangular correlation function, Eq.~\eqref{eq:corr-ricehat}. In both cases, the noise variance is $\sigma_n^2 = 0.1$.

As expected, the features in the power spectra that are discouraged by the spectral smoothness prior are not well reconstructed. From Fig.~\ref{fig:powspec-problematic} it is obvious that in the case with a Gaussian correlation function, the reconstructed and true power spectra deviate quite strongly on small scales. While the true power spectrum drops off rapidly, the reconstructed one stays comparatively flat. In the case of a triangular correlation function, the same can be said about the finite $k$-value where the true power spectrum drops to zero. While the drop is indeed represented by the signal realization, as can be seen from the dashed line in the lower panel, the reconstructed power spectrum stays relatively level.

However, the features in the power spectra that the filter fails to reconstruct are below the noise level, whereas the rise of the power spectrum in the bottom panel of Fig.~\ref{fig:powspec-problematic} on the smallest scales, which is above the noise level, is indeed represented in the reconstructed power spectrum as well. Thus, the signal reconstruction does not suffer significantly, as can be seen in Fig.~\ref{fig:maps-problematic}. An accurate reconstruction of the power spectrum is important mainly at the scales on which the signal-response and noise are of comparable magnitude. If the signal power is dominant, the reconstruction will follow the data closely, while it will smooth the data heavily in the opposite case of dominating noise, irrespective of the exact shape of the power spectrum. Thus, the reconstruction algorithm will in general perform well even in cases in which certain features in the power spectra are not allowed by the spectral smoothness prior. However, if the objective is an accurate reconstruction of the power spectrum and any such features are suspected to be present, the spectral smoothness prior needs to be adapted to this situation. Depending on the situation, the presence of these features is either expected a priori, e.g.\ in the case of spectral lines, or is manifest in the data, in which case they will be reconstructed even with a mild spectral smoothness prior, such as the features above the noise level in Fig.~\ref{fig:powspec-problematic}. If the presence of non-smooth spectral features is neither known a priori, nor contained to a significant degree in the data, a reconstruction of these features cannot be expected, independent of whether or not a spectral smoothness prior is used.

\section{Reconstructing log-normal fields}
\label{sec:log-normal}

Now we turn to the problem of reconstructing a log-normal field. We define our signal field $s$ to be the logarithm of this log-normal field, so that the prior probability distribution for $s$ is again a Gaussian, described by a mean and a covariance $S$. For simplicity, we assume the prior mean to be zero. The data model then becomes
\begin{equation}
    d = R \mathrm{e}^s + n,
\end{equation}
where we have again included additive Gaussian noise $n$ and the application of the exponential function to the signal field is to be interpreted pointwise. Due to the non-linearity of the exponential function, the posterior of $s$,
\begin{equation}
    \mathcal{P}(s|d,S) \propto \mathcal{G}{\left(d - R\mathrm{e}^s, N\right)}\, \mathcal{G}{\left(s,S\right)},
\end{equation}
is highly non-Gaussian, even when the signal covariance $S$ is known. Adding a prior for $S$ and marginalizing over it only makes the problem more complex,
\begin{equation}
    \mathcal{P}(s|d) \propto \mathcal{G}{\left(d - R\mathrm{e}^s, N\right)}\, \int \!\! \mathcal{D}S\, \mathcal{G}{\left(s,S\right)}\, \mathcal{P}(S).
\end{equation}

The path we pursue here is to treat this probability distribution approximatively. In order to do so, we employ the formalism of minimum Gibbs free energy presented in \cite{ensslin_weig-2010}. The basic idea is to approximate the posterior $\mathcal{P}(s|d)$ with a Gaussian, described by a mean $m$ and a covariance $D$. An approximate Gibbs free energy can be calculated as a function of these quantities and it was shown in \cite{ensslin_weig-2010} that the optimal Gaussian approximation -- in the sense of minimum Kullback-Leibler divergence -- can be obtained by minimizing the approximate Gibbs free energy with respect to $m$ and $D$.

The approximate Gibbs free energy is defined as
\begin{equation}
    \tilde{G}(m,D) = \tilde{U}(m,D) - T S_\mathrm{B}(D).
\end{equation}
The last term in this definition is the Boltzmann entropy
\begin{equation}
    S_\mathrm{B}(D) = \frac{1}{2} \mathrm{tr}{\left(1 + \log{\left(2\pi D\right)}\right)},
\end{equation}
which depends only on the covariance $D$. The other term in the approximate Gibbs free energy is the approximate internal energy
\begin{equation}
    \tilde{U}(m,D) = \left< H(s,d) \right>_{\mathcal{G}(s-m,D)}.
\end{equation}
Here, $\left<\cdot\right>_{\mathcal{G}(s-m,D)}$ denotes an expectation value calculated with respect to the Gaussian posterior approximation and $H(s,d) = -\log \mathcal{P}(s,d)$ is the Hamiltonian of the problem. Finally, the temperature $T$ can be regarded as a tuning parameter that regulates the importance that is given to the region around the maximum of the full posterior and regions removed from the maximum. The limiting value $T = 0$ leads to the maximum a posteriori solution for the signal field. In the following, we choose the default value of $T = 1$ and refer the reader to \cite{ensslin_weig-2010,iatsenko-2012,ensslin_weig-2012} for an in-depth discussion of the temperature parameter.

Assuming independent inverse-gamma distributions as priors for the power spectrum components $P_k$, as we had already done in Eq.~\eqref{eq:inverse-gamma}, the calculation of the Hamiltonian yields
\begin{align}
    H(s,d) &= -\log{\left( \mathcal{P}(d|s)\, \mathcal{P}(s|P)\, \mathcal{P}(P)\right)}\nonumber\\
    &= -\log{\left( \mathcal{G}{\left(d - R\mathrm{e}^s,N\right)}\, \mathcal{G}(s,S) \prod_k \mathcal{P}_\mathrm{IG}{(P_k)} \right)}\nonumber\\
    &= -j^\dagger \mathrm{e}^s + \frac{1}{2} \left(\mathrm{e}^s\right)^\dagger M \mathrm{e}^s\nonumber\\
    &\quad + \sum_k \left(\alpha_k - 1 + \frac{\rho_k}{2}\right) \log{\left( q_k + \frac{1}{2} \sum_{\left\{\vec{k}'|k' = k\right\}} \!\!\!\!\! \left|s_{\vec{k}'}\right|^2 \right)}\nonumber\\
    &\quad + \mathrm{const.}
\end{align}
Here, we have again collected $s$-independent terms in an additive constant and introduced the abbreviation
\begin{equation}
    M = R^\dagger N^{-1} R.
\end{equation}
In order to calculate the Gaussian expectation value of this Hamiltonian analytically, we expand the logarithm appearing in this expression in a power series around its expectation value
\begin{align}
    \tilde{q}_k &= \left< q_k + \frac{1}{2} \sum_{\left\{\vec{k}'|k' = k\right\}} \!\!\!\!\! \left|s_{\vec{k}'}\right|^2 \right>_{\mathcal{G}(s-m,D)}\nonumber\\
    &= q_k + \frac{1}{2} \sum_{\left\{\vec{k}'|k' = k\right\}} \!\!\!\!\! \left(\left|m_{\vec{k}'}\right|^2 + D_{\vec{k}'\vec{k}'}  \right)
\end{align}
as
\begin{align}
    &\log{\left( q_k + \frac{1}{2} \sum_{\left\{\vec{k}'|k' = k\right\}} \!\!\!\!\! \left|s_{\vec{k}'}\right|^2 \right)}\nonumber\\
    &\quad \approx \log{\left(\tilde{q}_k\right)}\nonumber\\
    &\quad\quad - \sum_{i=1}^{i_\mathrm{max}} \frac{(-1)^i}{i \tilde{q}_k^i} \left( q_k + \frac{1}{2} \left(\sum_{\left\{\vec{k}'|k' = k\right\}} \!\!\!\!\! \left|s_{\vec{k}'}\right|^2 \right) - \tilde{q}_k\right)^i.
\end{align}
We truncate this expansion after the first order, i.e.\ $i_\mathrm{max} = 1$. Note that our choice of $\tilde{q}_k$ ensures that the first-order term itself vanishes.

With this simplification, we can calculate the approximate Gibbs free energy to be
\begin{align}
    \tilde{G}(m,D) &= \int_\mathcal{M} \!\!\! \mathrm{d}x\, j_x \mathrm{e}^{m_x + \frac{1}{2} D_{xx}}\nonumber\\
    &\quad + \int_\mathcal{M} \!\!\! \mathrm{d}x \int_\mathcal{M} \!\!\! \mathrm{d}y\, \frac{1}{2} M_{xy} \mathrm{e}^{m_x + m_y + \frac{1}{2} D_{xx} + \frac{1}{2} D_{yy} + D_{xy}}\nonumber\\
    &\quad + \sum_k \left( \alpha_k - 1 + \frac{\rho_k}{2} \right) \log{\left(\tilde{q}_k\right)}\nonumber\\
    &\quad - \frac{1}{2} \mathrm{tr}{\left( 1 + \log{\left( 2\pi D\right)}\right)}.
\end{align}
To avoid confusion, we write out explicitly all integrals appearing here and in the following. Taking the functional derivatives with respect to $m$ and $D$ and equating them with zero yields two filter equations that determine $m$ and $D$. In these equations, the right hand side expression of Eq.~\eqref{eq:critical-P} appears. If we reidentify this expression with the spectral components $P_k$ and write $S_{\vec{k}\vec{k}'} = \delta_{\vec{k}\vec{k}'} P_k$, the two filter equations become
\begin{align}
    m_x &= \int_\mathcal{M} \!\!\! \mathrm{d}y\, S_{xy} \left[\vphantom{\int_\mathcal{M}} j_y \mathrm{e}^{m_y + \frac{1}{2}D_{yy}}\right.\nonumber\\
    &\quad - \left.\int_\mathcal{M} \!\!\! \mathrm{d}z\, M_{yz} \mathrm{e}^{m_y + m_z + \frac{1}{2} D_{yy} + \frac{1}{2} D_{zz} + D_{yz}} \right]
    \label{eq:log-normal-m}
\end{align}
and
\begin{align}
    \left(D^{-1}\right)_{xy} &= -j_x \mathrm{e}^{m_x + \frac{1}{2} D_{xx}} \delta_{xy}\nonumber\\
    &\quad + \int_\mathcal{M} \!\!\! \mathrm{d}z\, M_{xz} \mathrm{e}^{m_x + m_z + \frac{1}{2}D_{xx} + \frac{1}{2}D_{zz} + D_{xz}} \delta_{xy}\nonumber\\
    &\quad + M_{xy} \mathrm{e}^{m_x + m_y + \frac{1}{2} D_{xx} + \frac{1}{2} D_{yy} + D_{xy}}\nonumber\\
    &\quad + \left(S^{-1}\right)_{xy}.
    \label{eq:log-normal-D}
\end{align}

Together with Eq.~\eqref{eq:critical-P}, the last two equations fully determine the Gaussian approximation to the posterior. Solving them self-consistently gives an estimate $m$ for the signal field and an estimate $D$ for the corresponding uncertainty matrix. The corresponding approximate posterior mean estimate of the exponentiated field values then is
\begin{equation}
    \left< \mathrm{e}^{s_x} \right>_{\mathcal{P}(s|d)} \approx \left< \mathrm{e}^{s_x} \right>_{\mathcal{G}(s-m,D)} = \mathrm{e}^{m_x + \frac{1}{2}D_{xx}}.
\end{equation}

Of special interest for many applications is the case in which the matrix $M = R^\dagger N^{-1} R$ is diagonal, e.g.\ when the noise contributions to the individual data points are uncorrelated and the response is purely local. In this case the filter equations simplify somewhat to
\begin{equation}
    m_x = \int_\mathcal{M} \!\!\! \mathrm{d}y\, S_{xy} \left[ j_y \mathrm{e}^{m_y + \frac{1}{2} D_{yy}} - M_{yy} \mathrm{e}^{2 m_y + 2 D_{yy}} \right]
    \label{eq:log-normal-m-simplistic}
\end{equation}
and
\begin{align}
    \left(D^{-1}\right)_{xy} &= \left[ -j_x \mathrm{e}^{m_x + \frac{1}{2} D_{xx}} + 2 M_{xx} \mathrm{e}^{2 m_x + 2 D_{xx}} \right] \delta_{xy}\nonumber\\
    &\quad + \left(S^{-1}\right)_{xy}.
    \label{eq:log-normal-D-simplistic}
\end{align}

\subsection{Spectral smoothness in the log-normal case}

In Sec.~\ref{sec:combi-derivation} we had seen that in the Gaussian case the full reconstruction with unknown power spectrum can be regarded as a combination of the posterior mean reconstruction under the assumption of a known power spectrum and the estimation of this power spectrum as the one that maximizes its posterior probability. In the case of a log-normal field, the posterior mean cannot be calculated analytically, even if the power spectrum is assumed to be known. However, it is interesting to note that employing the formalism of minimum Gibbs free energy to the log-normal reconstruction problem with known power spectrum, one arrives exactly at the formulas derived in the previous section, Eqs.~\eqref{eq:log-normal-m} and \eqref{eq:log-normal-D}.

Thus, the full reconstruction, consisting of Eqs.~\eqref{eq:log-normal-m}, \eqref{eq:log-normal-D}, and \eqref{eq:critical-P} can again be regarded as a combination of the calculation of the posterior mean for the signal under the assumption of a power spectrum and the estimation of the power spectrum according to Eq.~\eqref{eq:critical-P}. From this viewpoint, the inclusion of a smoothness prior for the power spectrum is trivial. We simply replace the power spectrum estimation step according to Eq.~\eqref{eq:critical-P} with the one derived in Sec.~\ref{sec:smoothness-prior}, i.e.\ with Eq.~\eqref{eq:critical-P-withsmooth}, just as we had done in the Gaussian case.

\subsection{Test cases}
\label{sec:lognormal-tests}

\begin{figure}[!]
    \input{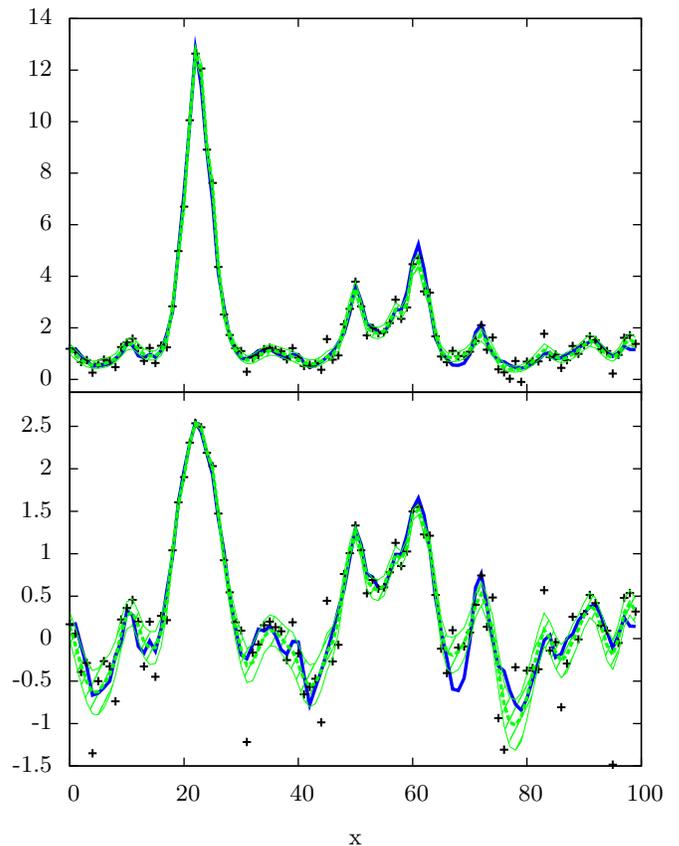}
    \caption{\label{fig:linint}One-dimensional log-normal reconstruction in a mildly non-linear low-noise case. The power spectrum from which the signal realization was drawn is given by Eq.~\eqref{eq:powspec} with $P_0 = 0.1$, $k_0 = 5$, and $\gamma = 4$. The noise variance is $\sigma_n^2 = 0.1$. The top panel shows the exponentiated signal field $\mathrm{e}^s$ (blue solid line), the data $d$ (crosses), the reconstruction $\mathrm{e}^{m + \frac{1}{2}\mathrm{diag}(D)}$ (green dashed line), and the uncertainty interval of the reconstruction, given by $\mathrm{e}^{m + \frac{1}{2}\mathrm{diag}(D) \pm (\mathrm{diag}(D))^{1/2}}$ (hatched region). The lower panel shows the signal field $s$ (blue solid line), the logarithm of the data $\log (d)$ (crosses), the reconstruction $m$ (green dashed line), and the uncertainty interval for $m$, given by $m \pm (\mathrm{diag}(D))^{1/2}$. In the lower panel, only data points for which $\log d$ is greater than $-1.5$ are shown.}
\end{figure}

\begin{figure}[!]
    \input{map-noninta.tex}
    \caption{\label{fig:nonint}One-dimensional log-normal reconstruction in a highly non-linear low-noise case. The plotted quantities are the same as in Fig.~\ref{fig:linint}. The parameters describing the power spectrum are $P_0 = 0.3$, $k_0 = 5$, and $\gamma = 4$ and the noise variance is $\sigma_n^2 = 1$. In the lower panel, only data points for which $\log d$ is greater than $-3$ are shown.}
\end{figure}

\begin{figure}[!]
    \input{map-nonhigha.tex}
    \caption{\label{fig:nonhigh}One-dimensional log-normal reconstruction in a highly non-linear high-noise case. The plotted quantities are the same as in Fig.~\ref{fig:linint}. The parameters describing the power spectrum are $P_0 = 0.3$, $k_0 = 5$, and $\gamma = 4$ and the noise variance is $\sigma_n^2 = 25$. In the lower panel, only data points for which $\log d$ is greater than $-3$ are shown.}
\end{figure}

\begin{figure*}
    \includegraphics[width=\textwidth]{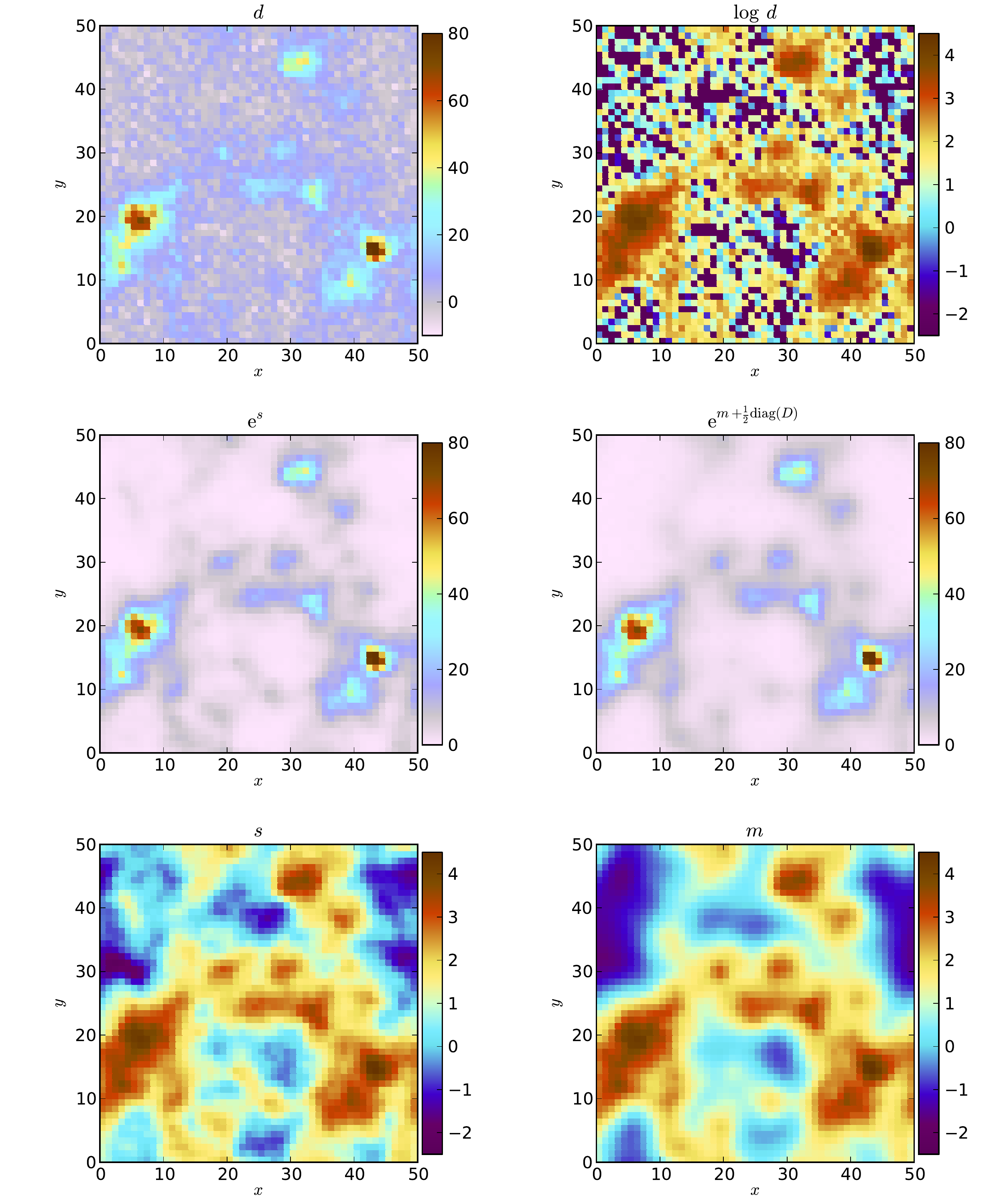}
    \caption{\label{fig:toroidal}Log-normal reconstruction example in a toroidal setting. The signal's power spectrum is given by Eq.~\eqref{eq:powspec} with $P_0 = 0.3$, $k_0 = 2$, and $\gamma = 5$. The noise variance is $\sigma_n^2 = 10$. The top row shows the data set $d$ (left) and its logarithm (right). In the logarithmic version, pixels with negative data values are plotted as dark blue. The middle row shows the exponentiated signal field $\mathrm{e}^s$ (left) and its reconstruction, given by $\mathrm{e}^{m + \frac{1}{2}\mathrm{diag}(D)}$, (right). The corresponding non-exponentiated quantities $s$ (left) and $m$ (right) are shown in the bottom row.}
\end{figure*}

\begin{figure*}
    \includegraphics[width=\textwidth]{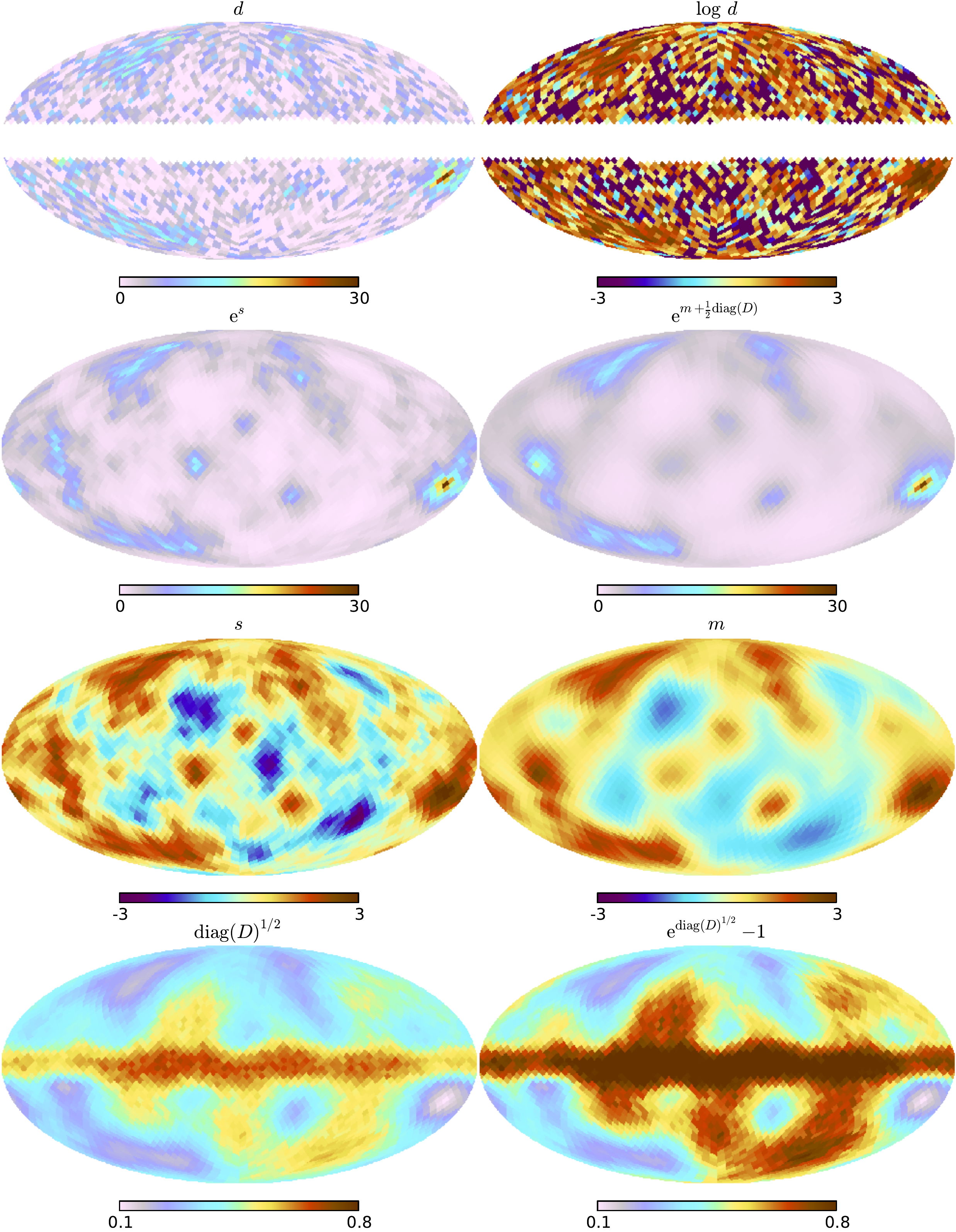}
    \caption{\label{fig:spherical}Log-normal reconstruction example on a two-sphere. The quantities plotted in the top three rows are the same as in Fig.~\ref{fig:toroidal}. The bottom row shows the uncertainty of the signal reconstruction given by $\mathrm{diag}{(D)}^{1/2}$ (left side) and the fractional uncertainty of the reconstruction of the exponentiated field given by $\mathrm{e}^{\mathrm{diag}{(D)}^{1/2}} - 1$ (right side). The power spectrum parameters are $P_0 = 0.3$, $k_0 = 5$, and $\gamma = 4$ and the noise variance is $\sigma_n^2 = 10$. In the top row, pixels without measurements are plotted white.}
\end{figure*}

In this section, we present some test cases for the theory developed so far on the reconstruction of log-normal fields. We study one-dimensional tests with differing degrees of non-linearity and differing noise-levels, as well as two two-dimensional test-cases. For simplicity, we assume an ideal local response, $R = \mathbb{1}$, except in the last example, where we study the effects of an observational mask. The noise is assumed to be uncorrelated and homogeneous, $N = \sigma_n^2 \mathbb{1}$. For the smoothness prior, we use the one discussed in Sec.~\ref{sec:smoothness-prior} and choose $\sigma_p^2 = 100$.

First, we discuss again the case of a signal on the one-sphere $\mathcal{S}^1$, i.e.\ the interval $[0,1)$ with periodic boundary conditions, here discretized into $100$ pixels. Shown in Figs.~\ref{fig:linint}-\ref{fig:nonhigh} are three different cases. In each of these cases, the signal realization is the same as in Sec.~\ref{sec:gaussian}, only with a differing normalization of the power spectrum. In Fig.~\ref{fig:linint} we show a mildly non-linear case with $P_0 = 0.1$ and in Figs.~\ref{fig:nonint} and \ref{fig:nonhigh} a highly non-linear case with $P_0 = 0.3$. The noise variance $\sigma_n^2$ is $0.1$ in Fig.~\ref{fig:linint}, $1$ in Fig.~\ref{fig:nonint}, and $25$ in Fig.~\ref{fig:nonhigh}. In the lower panels of these three figures, we show the signals and their reconstructions and in the upper panels the exponentiated fields, $\mathrm{e}^s$, as well as our posterior mean estimate for these, $\left<\mathrm{e}^s\right>_{\mathcal{P}(s|d)} \approx \mathrm{e}^{m + \frac{1}{2}\mathrm{diag}(D)}$.

In the mildly non-linear case the reconstruction is a reasonably good approximation to the true signal both in regions of high signal values and regions of low signal values. However, it is apparent that the quality of the reconstruction is slightly higher in the former regions. This is due to the homogeneity of the noise statistics that we have assumed. Since our signal response, given by $R\mathrm{e}^s$, depends non-linearly on the signal, the noise impact is lower in regions of higher signal values and hence the signal inference is less demanding in these regions. From Fig.~\ref{fig:linint} it is apparent that this effect is well represented by the point-wise uncertainty of the reconstruction, given by $\mathrm{diag}(D)$. As can be seen from Fig.~\ref{fig:nonint}, the effect becomes more pronounced for higher degrees of non-linearity, i.e.\ larger signal variances on a logarithmic scale.

The highly nonlinear case with high noise level, depicted in Fig.~\ref{fig:nonhigh}, exhibits a point-wise uncertainty estimate for the reconstruction that can clearly not be interpreted as a 68\% confidence interval. This is due to a known problem of the filter discussed in Sec.~\ref{sec:gaussian} and by extension also of the filter derived in this section. As discussed in \cite{ensslin_frommert-2011}, the filter exhibits a perception threshold. This means that if the signal-response-to-noise ratio is lower than a certain threshold on a given scale, then the filter will not reconstruct any power on this scale. Our usage of the spectral smoothness prior partly alleviates this porblem in that it prevents the power of individual scales to drop to zero. However, the reconstructed power on the noise-dominated scales will in general still be too low. This directly affects the estimate for the reconstruction's uncertainty, given by $D = \left(S^{-1} + M\right)^{-1}$, which tends toward zero in the limit of zero power on all scales.

Furthermore, the lack of power on all but the few signal-dominated scales can lead to ringing effects, i.e.\ prominent signal-dominated features are well reconstructed and extrapolated periodically into the noise-dominated regions. The deep trough in the signal reconstruction that can be seen in the lower panel of Fig.~\ref{fig:nonhigh} around pixel number $35$ is most likely due to this effect. The high degree of non-linearity acts to reinforce this effect. As can be seen in the top panel of Fig.~\ref{fig:nonhigh}, the effect of the trough onto the exponentiated reconstruction is almost intangible.

It is certainly mandatory to keep the potential problems of ringing and an underestimated uncertainty in mind when reconstructing a field which is swamped by noise in the better part of its domain. However, as can be seen from Fig.~\ref{fig:nonhigh}, the main features of the signal field will still be reconstructed reliably even in such an unfavorable case.

In these test cases, we have chosen the one-dimensional interval as domain of the log-normal signal field mainly for illustrative purposes. The algorithm is, however, versatile in that it can operate on virtually any space\footnote{We are using the \textsc{nifty} package \citep{selig-2013} in our implementation. This makes changing the domain of the signal field trivial, requiring only minuscule changes to the code.}. To illustrate this point, we close this section with two two-dimensional examples.

In Fig.~\ref{fig:toroidal}, we show an example for a signal defined on the two-torus $\mathcal{T}^2 = \mathcal{S}^1 \times \mathcal{S}^1$, i.e.\ a segment of $\mathbb{R}^2$ with periodic boundary conditions. The periodicity here and in the earlier one-dimensional examples is a consequence of our use of Fast Fourier Transform routines. It can be approximately overcome by artificially extending the interval or the section of $\mathbb{R}^2$ for more than the typical correlation length of the signal while leaving the data unchanged when doing the reconstruction. Thus, the influence of the reconstruction on one side of the interval or rectangle will have negligible influence on the reconstruction on the other side.

We discretize the two-torus and its corresponding Fourier space into $50 \times 50$ pixels. The rectangular orientation of the Fourier pixels leads to many different scales $k$ being represented, each, however, only by a few pixels, corresponding to a few different Fourier vectors $\vec{k}$. To avoid dealing with all these scales individually, we bin the scales logarithmically into $17$ bins. We replace all summations over Fourier components with a certain scale appearing in the filter formulas with summations over all Fourier components whose scale falls within one bin. Thus, we reconstruct the power for each bin instead of for each scale that is represented in the rectangular Fourier grid. The signal is again drawn from the power spectrum given by Eq.~\eqref{eq:powspec} with $P_0 = 0.3$, $k_0 = 2$, and $\gamma = 5$. The noise level is chosen as $\sigma_n^2 = 10$. Signal, data, and reconstruction are shown in Fig.~\ref{fig:toroidal}.

The toroidal example exhibits essentially the same features that we had already seen in the one-dimensional case. While the reconstruction is generally in good agreement with the true underlying signal, it is less accurate in the regions of small signal values where the signal-response-to-noise ratio is far worse due to the quite high degree of non-linearity.

In Fig.~\ref{fig:spherical}, we show an example for a reconstruction on the two-sphere $\mathcal{S}^2$. This example has special relevance for astronomical applications since one can interpret the sky as a two-sphere and therefore any astrophysical signals without distance information will be defined on this manifold. We generate a mock signal from the power spectrum given in Eq.~\eqref{eq:powspec} with $P_0 = 0.3$, $k_0 = 5$, and $\gamma = 4$. The noise level is $\sigma_n^2 = 10$. We discretize the sphere using the \textsc{HEALPix}\footnote{The \textsc{HEALPix} package is available from \url{http://healpix.jpl.nasa.gov/}.} package \citep{gorski-2005} with a resolution parameter of $N_\mathrm{side} = 16$ leading to $3\,072$ pixels in total. The power spectrum components $P_k$ are in this case the components of an angular power spectrum, often denoted as $C_\ell$.

In this last case study, we replace the trivial response $R = \mathbb{1}$ with a projection onto part of the sphere, effectively masking $600$ pixels around the equator for which we assume that no measurements have been taken. This resembles a typical situation in extragalactic astronomy, where observations through the Galactic plane are not possible due to the obscuration by the Milky Way.

Fig.~\ref{fig:spherical} shows the signal field, data, and reconstruction both in the exponentiated and in the linear version. In the panels showing the reconstructed signal field, it can be nicely seen how the algorithm is able to extrapolate into the gap region from the data on the boundary. This is possible due to the knowledge of the correlation structure that was inferred from the same data set. Also shown in Fig.~\ref{fig:spherical} are the pixel-wise uncertainty estimate of the signal field's reconstruction, given by $\mathrm{diag}{(D)}^{1/2}$, and the fractional uncertainty of the reconstructed exponentiated signal, given by $\mathrm{e}^{\mathrm{diag}{(D)}^{1/2}} - 1$, which is approximated well by $\mathrm{diag}{(D)}^{1/2}$ in most regions. It can be seen that the uncertainty tends to be higher in the regions of low signal values, as was the case in the one-dimensional examples, and in the region around the equator which lacks observations. This is to be expected, since the only constraint on the signal in this region comes from extrapolations from neighboring regions using the signal's correlation structure inferred from the data.

\section{Summary and Conclusions}
\label{sec:conclusions}

We have developed an algorithm to infer log-normal random fields from noisy measurement data. The log-normal model was chosen due to its wide range of applications in astrophysics and other fields. The reconstruction method uses the correlation structure of the log-normal field to differentiate between features in the data that are due to noise and such that are due to variations in the true underlying field. This correlation structure, determined by the field's power spectrum, is, however, in general not known a priori. We have therefore extended the theory for simultaneous reconstructions of a field and its power spectrum that was developed and applied successfully for Gaussian random fields in the past \citep{ensslin_frommert-2011, ensslin_weig-2010, oppermann-2011a, oppermann-2011b, oppermann-2012a, selig-2012} to log-normal fields.

An additional feature of our reconstruction method is the use of a smoothness prior for the power spectrum. We have suggested to employ a prior based on the second double-logarithmic derivative of the power spectrum and shown that it is well suited to handle a large variety of cases. A beneficial feature of this particular approach is that it only adds one simple term to the formula for the power spectrum estimation and is therefore easy to implement. Having investigated possible pitfalls associated with the usage of such a prior from theoretical as well as practical viewpoints, we should stress that the derivation of the filter formulas laid out here does not depend on the specific form of the spectral smoothness prior. In cases in which the prior we employed here cannot be expected to yield satisfactory results it should simply be replaced by a different one.

The algorithm we have derived depends in no way on the space on which the signal field is to be reconstructed. We have demonstrated this by showing examples of reconstructions of mock signals on a one-dimensional interval, a flat two-dimensional space, and a spherical space. We have discussed the performance of the algorithm in these scenarios and pointed out possible caveats when dealing with very low signal-response-to-noise ratios.

In these application examples, we have assumed that the observational data simply represent the underlying log-normal field, subject to additive noise. Furthermore, we have illustrated the ability of the algorithm to extrapolate from the given data in a test case in which these data were assumed to be incomplete, thus demonstrating the power of the usage of the correlation information contained in the data. The derivation of the filter formulas, however, is even more general. It allows for an arbitrary linear relationship between log-normal field and data, described by a response matrix $R$. The resulting formulas include the general response matrix which can e.g.\ represent an incomplete observation, a convolution, or a Fourier transformation. We have also allowed for a general noise covariance matrix, thus including cases of heteroscedastic or correlated noise.

This makes the algorithm widely applicable. Applications that we have in mind include for example the study of diffuse Galactic emission components at radio frequencies and reconstructions of emissivity fields across galaxy clusters from interferometric observations.

\begin{acknowledgments}
    The authors would like to thank Maksim Greiner and Henrik Junklewitz for enlightening discussions, as well as an anonymous referee and Benjamin D.\ Wandelt for helpful comments. Some of the results in this paper have been derived using the \textsc{HEALPix} \citep{gorski-2005} package. The calculations were performed using the \textsc{nifty} package \citep{selig-2013}. This research has made use of NASA's Astrophysics Data System. MRB is supported by the DFG Forschergruppe 1254 Magnetisation of Interstellar and Intergalactic Media: The Prospects of Low-Frequency Radio Observations.
\end{acknowledgments}

\bibliographystyle{myaa}
\bibliography{lnsmooth}

\appendix

\section{Case studies for the spectral smoothness prior}
\label{app:casestudies}

In this appendix, we discuss a few possible correlation structures and power spectra and examine the suitability of a spectral smoothness prior of the form described by Eq.~\eqref{eq:smoothnessprior}.

\subsection{Power law spectra}
\label{app:powerlaws}

If the signal's power spectrum is a broken power law of the shape given by Eq.~\eqref{eq:powspec}, the first double-logarithmic derivative is
\begin{equation}
    \frac{\partial p_k}{\partial \log k} = -\frac{\gamma}{1 + \left(\frac{k_0}{k}\right)^2},
\end{equation}
which behaves like $-\gamma\left(\frac{k}{k_0}\right)^2$ for $k \ll k_0$ and tends toward $-\gamma$ for $k \rightarrow \infty$. The second double-logarithmic derivative is
\begin{equation}
    \frac{\partial^2 p_k}{\partial \left(\log k\right)^2} = -\frac{2\gamma}{\left( \frac{k}{k_0} + \frac{k_0}{k} \right)^2},
\end{equation}
which tends toward zero both for $k \rightarrow 0$ and $k \rightarrow \infty$. The second derivative takes on its extremum, given by $-\frac{\gamma}{2}$, at the knee frequency $k_0$ where the spectral index of the power spectrum changes from $0$ to $\gamma$. Thus neither a prior punishing large values for the first logarithmic derivative, nor one punishing large values for the second logarithmic derivative prevents the true solution from being found, provided the values of $-\gamma$ and $-\frac{\gamma}{2}$, respectively, are well within the range of values that are allowed for the derivative by the prior. In fact, choosing $\sigma_p = \gamma/2$ would allow such a change in spectral index roughly once per e-folding of the $k$-value. Choosing $\sigma_p = \frac{\gamma}{2h}$ would turn such a kink into a less common $h$-sigma event.

The case of a broken power-law contains the special cases of a pure power-law, i.e.\ $k_0 \rightarrow 0$, such as arises for example for Brownian motion of a particle, and of an exponential two-point correlation function, i.e.
\begin{equation}
    C(r) = S_{xy} = C_0 \mathrm{e}^{-\beta r},
\end{equation}
where $r = \left|x - y\right|$ is the distance between two points or time-instances $x$ and $y$. Such a correlation function arises for example from the Ornstein-Uhlenbeck process \citep{uhlenbeck-1930}. Fourier transforming this correlation function in a one-dimensional space yields the power spectrum which takes on the form given by Eq.~\eqref{eq:powspec} with $P_0 = \frac{2 C_0}{\beta}$, $k_0 = \beta$, and $\gamma = 2$.

However, there are also cases that do not lead exclusively to small values of the double-logarithmic derivatives. Two such scenarios will be studied in the remainder of this appendix.

\subsection{Gaussian correlations}
\label{app:Gaussian}

Consider a two-point correlation function of Gaussian shape for a field on a one-dimensional space, i.e.
\begin{equation}
    C(r) = C_0 \mathrm{e}^{-\frac{r^2}{2\sigma^2}}.
    \label{eq:corr-gauss}
\end{equation}
A field whose statistics are described by such a two-point correlation function can be regarded as a stationary and spatially uncorrelated field convolved with a Gaussian. Calculating the corresponding power spectrum via Fourier transformation yields
\begin{equation}
    P_k = \sqrt{2\pi}\sigma C_0 \mathrm{e}^{-\frac{\sigma^2 k^2}{2}}.
\end{equation}
This power spectrum drops quickly with increasing $k$, due to the flatness of the correlation function around $r = 0$. Therefore, both the first and second double-logarithmic derivatives grow unbounded as $k \rightarrow \infty$. Thus, by employing a spectral smoothness prior that punishes large values for these derivatives, one prevents in principle the reconstruction of the true power spectrum and suppresses the small-scale correlations in the reconstructed field, which are in reality more pronounced. In cases in which Gaussian correlations are expected, it might therefore be advisable to choose the strength of the spectral smoothness prior, given by $\sigma_p$, $k$-dependent or choose a different smoothness prior altogether.

\subsection{Triangular correlations}
\label{app:ricehat}

Another case in which the double-logarithmic derivatives of the power spectrum can become divergent is a correlation function with finite support. As an example, we consider a signal on a one-dimensional space with correlations only over the finite distance $2L$, given by
\begin{equation}
    C(r) = \left\{\begin{array}{cc}
        C_0 \left(1 - \frac{r}{L}\right) & \textnormal{for}~ r < L\\
        0 & \textnormal{else}
    \end{array}\right..
    \label{eq:corr-ricehat}
\end{equation}
This triangular correlation function describes a stationary and spatially uncorrelated field that has been convolved with a top-hat kernel. Fourier transforming it yields the power spectrum
\begin{equation}
    P_k = \frac{2 C_0}{k^2 L} \left( 1 - \cos{\left(kL\right)} \right).
\end{equation}
This power spectrum becomes exactly zero at finite $k$-values. Its logarithm, and therefore also the double-logarithmic derivatives, are divergent at these locations, so not even a $k$-dependent value of $\sigma_p$ can ensure the correct reconstruction of the power spectrum in this case. The oscillatory behavior of this power spectrum is a generic feature of signal fields that are correlated only over a finite distance.

Note that both in the case of only locally correlated fields and in the case of Gaussian correlations, the double-logarithmic derivatives can be kept finite by adding a constant floor to the power spectrum, i.e.\ introducing an additive part to the signal that is spatially uncorrelated. In many cases, the variance of the uncorrelated addition needed to satisfy the spectral smoothness prior, Eq.~\eqref{eq:smoothnessprior}, will be small enough so as not to influence the signal reconstruction significantly. Note that in practice, any field variations are restricted by the finite pixel size and only $k$-values up to a finite $k_\mathrm{max}$ will be considered. In Sec.~\ref{sec:gaussian-test-cases}, we investigate the effect that the spectral smoothness prior given in Eq.~\eqref{eq:smoothnessprior} has in practice on the recunstruction of a signal field exhibiting the two potentially problematic two-point correlations that we discussed here.

In conclusion, using a spectral smoothness prior that punishes large values for the first or second double-logarithmic derivative of the power spectrum can lead to the introduction of spurious small-scale variations in cases in which the true two-point correlation function is flat around $r=0$ or has only finite support. In the latter case, it can also introduce spurious large-scale correlations. This will, however, in general only be a problem for the reconstruction if some feature that mimics these large-scale correlations is present in the data, i.e.\ caused by the noise. Another case in which the employment of any spectral smoothness prior is obviously a bad idea is that of a signal that exhibits prominent periodicities. The detection of such spectral lines would only get hindered by the usage of a spectral smoothness prior.

\section{Discretization of the spectral smoothness prior}
\label{app:T-matrix}

In our implementation we use discretized values $\left(k_i\right)_{i=0,\dots,i_\mathrm{max}}$ for the length scales that are represented on the computational grid, or in case of the two-torus, bins thereof. We approximate the integral in the exponent of the spectral smoothness prior, Eq.~\eqref{eq:smoothnessprior}, with a sum according to
\begin{align}
    \int \!\! \mathrm{d}{\left(\log k\right)}\, \left(\frac{\partial^2 \log P_k}{\partial \left(\log k\right)^2}\right)^2 \approx & \sum_{i=1}^{i_\mathrm{max}-1} \delta_i \left( \Delta p \right)_i^2.
\end{align}
Here, we use the abbreviations
\begin{equation}
    \delta_i = \frac{\log k_{i+1} - \log k_{i-1}}{2}
\end{equation}
for integer values of $i$ and
\begin{equation}
    \delta_i = \log k_{i + 1/2} - \log k_{i - 1/2}
\end{equation}
for half-integer values of $i$. Note that we have excluded the boundaries at $k_0$ and $k_{i_{\mathrm{max}}}$ from the sum to avoid numerical problems at these locations.

We approximate the second logarithmic derivative as
\begin{equation}
    \left(\Delta p \right)_i = \sum_j \Delta_{i,j} p_j = \frac{\frac{p_{i+1} - p_i}{\delta_{i + 1/2}} - \frac{p_i - p_{i-1}}{\delta_{i - 1/2}}}{\delta_i},
\end{equation}
so that we can represent it as a matrix with the entries
\begin{align}
    \Delta_{i,i} &= -\frac{1}{\delta_i} \left( \frac{1}{\delta_{i + 1/2}} + \frac{1}{\delta_{i - 1/2}} \right),\\
    \Delta_{i,i \pm 1} &= \frac{1}{ \delta_i \delta_{i \pm 1}},
\end{align}
acting on the vector $p$. All other entries of the matrix $\Delta$ are zero.

We can now write the exponent of Eq.~\eqref{eq:smoothnessprior} as
\begin{equation}
    -\frac{1}{2\sigma_p^2}\int \!\! \mathrm{d}{\left(\log k\right)}\, \left(\frac{\partial^2 \log P_k}{\partial \left(\log k\right)^2}\right)^2 = -\frac{1}{2} \sum_{i,j = 0} p_i T_{i,j} p_j,
\end{equation}
where the matrix $T$ is given by
\begin{equation}
    T_{i,j} = \frac{1}{\sigma_p^2} \sum_l \Delta_{l,i} \delta_l \Delta_{l,j}.
\end{equation}

\end{document}